\documentstyle[aps,epsf,floats]{revtex}

\begin{document}
\title{Diffusion equation and spin drag in spin-polarized transport}
\author{Karsten Flensberg$^{1}$, Thomas Stibius Jensen$^{1}$, and Niels Asger
Mortensen$^{1,2}$}
\address{$^{1}${\O}rsted Laboratory, Niels Bohr Institute fAPG,
Universitetsparken 5, 2100 Copenhagen, Denmark. \\
$^{2}$Mikroelektronik Centret, Technical University of Denmark, 2800 Lyngby, Denmark.}
\date{\today}
\maketitle  \draft

\begin{abstract}
We study the role of electron-electron interactions for spin polarized transport using the
Boltzmann equation and derive a set of coupled transport equations. For spin polarized
transport the electron-electron interactions are important, because they tend to equilibrate
the momentum of the two spin species. This ``spin drag'' effect enhances the resistivity of
the system. The enhancement is stronger the lower the dimension and should be measurable in
for example a two dimensional electron gas with ferromagnetic contacts. We also include spin
flip scattering which has two effects: it equilibrates the spin density imbalance and,
provided it has a non $s$-wave component, also the current imbalance.
\end{abstract}

\section{Introduction}

Recent advances in fabrication of ferromagnetic--semiconductor heterostructures\cite{levy94}
and the observation of spin injection into semiconductors\cite{pasp2000} have lead to new
interest in transport properties of spin polarized systems. There has been considerable work
done in the field of metallic magnetic multilayers, which has been analyzed in terms of
transport equations with spin dependent distribution functions\cite{john87,son87}. These works
based their analysis on diffusion transport equations. The justification of using these
equations was given by Valet and Fert\cite{valetfert93}, who derived a spin diffusion
equation from the Boltzmann equation in the limit where the spin scattering length is much
longer than the momentum relaxation length. Recently, the transport equation were utilized to
analyze the feasibility of spin injection into semiconductors with the result that the
crucial parameter is the conductivity mismatch between the semiconductor and the
ferromagnet\cite{schm00} and to study theoretically spin-polarized transport in in
inhomogeneous doped semiconductors\cite{zutic01}.

None of these approaches take the electron-electron (e-e) scattering into account. Clearly
the e-e interactions play a different role than in usual spin degenerate transport, where the
e-e interaction does not provide a mechanism for momentum relaxation and hence has only
indirect consequences for transport coefficients. In spin polarized transport the two spin
species have different drift velocities and e-e interactions are instrumental in
equilibrating this difference. This leads to a spin drag effect where the spins carrying the
larger current will drag along the spins carrying the smaller current. This drag effect has
been considered recently by D'Amico and Vignale\cite{damicovignale00a,damicovignale00b} in
three dimensions using linear response theory. They found that the the spin drag resistivity
was appreciable and can at elevated temperatures become fractions of the usual resistivity of
the metal. In two dimensions the effect of e-e interactions on spin diffusion has been
considered theoretically by Takahashi et al.\cite{taka99,taka01}. Using a quantum kinetic
equation approach, previously utilized in $^{3}$He-$^{4}$He solutions\cite{mullinjeon92},
they studied the spin diffusion coefficients in two dimensional electron gases. In order to
study this spin diffusion they used variational functions but did not include spin relaxation
scattering.

In this paper, we use the Boltzmann equation to study spin dependent transport and spin
diffusion. We restrict our selves to the study of colinear magnetization and our goal is to
derive a set of transport equation in the semiclassical limit. For this purpose the Boltzmann
equation is adequate. For the non-colinear case, where phenomena like damped transverse spin
modes cannot occur, one must go beyond the present approach, see e.g. \cite{taka99,taka01} and
references therein. We include impurity scattering, both spin independent and spin flip
scattering, as well as e-e scattering. We show that a shifted Fermi--Dirac distribution
(SFD), is a valid solution at low temperatures, $T\ll T_{F}$ and without spin flip scattering.
This is also the case for weak e-e scattering, where the problem in absence of spin flip
reduces to the ordinary Coulomb drag situation\cite{jauhosmith93,flen95hu}.

We then go on to discuss the general case at higher temperatures, general interaction
strength and finite intrinsic spin flip scattering. Using a SFD ansatz, we find for an
isotropic system the following macroscopic transport equations
\begin{mathletters}
\label{macrotrans}
\begin{eqnarray}
\nabla \cdot {\mathbf{J}}_{s} &=&\left( -\frac{e}{\tau _{{\rm sf}}^{0}}\frac{\partial
n_{s}^{0}}{\partial \mu }\right) \left( \bar{\mu}_{s}-\bar{\mu}_{-s}\right) ,
\label{cont.eq} \\
\nabla \bar{\mu}_{s} &=&\frac{e}{\sigma _{s}}{\mathbf{J}}_{s}+ \left( \frac{e}{\sigma
_{D}\alpha _{s}}+\frac{e}{\sigma _{{\rm sf,s}}}\right) \left( {\mathbf{J}}_{s}-\alpha
_{s}{\mathbf{J}}_{-s}\right) . \label{ohmslaw}
\end{eqnarray}
Here $J_{s}$ is the current carried by electrons with spin $s$, $\bar{\mu}_{s}$ is the local
spin dependent electro chemical potential, $\sigma _{s}$ is the conductivity of the spin $s$
electron gas, $\tau _{{\rm sf}}^{0}$ is a spin lifetime due to intrinsic spin flip processes,
$\sigma _{{\rm sf,s}}$ is a spin current conversion conductivity arising from the angle
dependence of spin flip scattering, and $\alpha _{s}=n_{s}^{0}/n_{-s}^{0}$ is the relative
spin density, see Eqs.~(\ref{tausff}),(\ref{sigmas}) for definitions. Finally $\sigma _{D}$ is
the spin drag conductivity given by
\end{mathletters}
\begin{equation}
(\sigma _{D})^{-1}=\frac{\hbar ^{2}}{de^{2}n_{s}n_{-s}}\int \frac{d{\bf q}}
{(2\pi)^{d}}\int_{0}^{\infty }d\omega \,q^{2}|e\phi (q)|^{2}\frac{{\rm
\mathop{\rm Im}%
}\chi _{s}(q,\omega )%
\mathop{\rm Im}%
\chi _{-s}(q,\omega )}{k_{B}T\sinh ^{2}(\hbar \omega /2k_{B}T)}, \label{rhoD}
\end{equation}
where Im $\chi _{s}$ is the polarization function%
\begin{equation}
{\rm
\mathop{\rm Im}%
}\chi _{s}(q,\omega )=\pi \int \frac{d{\bf k}}{(2\pi )^{d}}\left[ f_{s}^{0}(|%
{\bf k+q|})-f_{s}^{0}(k)\right] \delta \left( \varepsilon _{|{\bf k+q}%
|s}-\varepsilon _{ks}-\hbar \omega \right) .
\end{equation}
The formula (\ref{rhoD}) is well-known from Coulomb drag\cite{jauhosmith93}. At low
temperatures it goes $(\sigma _{D})^{-1}\propto T^{2}$ in two and three dimensions, while in
one dimension it is proportional to $T$, see e.g. \cite{hu96flenb}.

The first transport equation (\ref{cont.eq}) is the continuity equation, which expresses the
conservation of current in the presence of spin flip processes. The second equation
~(\ref{ohmslaw}) is a generalized Ohm's law. The first term on the right hand side is the
usual Ohm's low, while the second term describes that a momentum imbalance between the two
spin directions gives rise to an additional resistance if there is a mechanism for conversion
of spin current. There are two such processes possible. This first one is the spin drag
effect mentioned above, where e-e scattering makes transfer of momentum possible. The second
one is due elastic spin flip scattering on for example magnetic impurities, which can convert
a current with one spin polarization to a current of the opposite polarization, if the spin
flip matrix element has an angular dependence. For example if the spin flip predominately
scatters forward it means that a spin flip scattering is accompanied with a transfer of
momentum. In contrast if the spin flip scattering is purely $s$-wave scattering the momentum
transfer between the spin channels is on average equal to zero. This can be seen
mathematically from the expression for $\tau _{{\rm sf}}$ in Eq.~(\ref{tausf}). The
derivation of the these two terms is the main result of the present paper.

Two consequences of the spin current relaxation terms can immediately be read off. Firstly,
they give rise to an increased resistivity in the case where the current is spin polarized.
For example, taking $J_{\downarrow }=0$ the effective resistivity for electrons with spin
$\uparrow $ becomes $(1/\sigma _{\uparrow }+1/\sigma _{D}\alpha _{s}+1/\sigma _{{\rm
sf,s}})^{-1}$ and hence an enhanced resistivity. Secondly, from Eqs.~(\ref{macrotrans}) we
get a diffusion equation for the electro chemical potential difference
\begin{equation}
\nabla ^{2}(\bar{\mu}_{s}-\bar{\mu}_{-s})=\frac{\bar{\mu}_{s}-\bar{\mu}_{-s}}{\ell _{{\rm
sf}}^{2}},
\end{equation}
where%
\begin{equation}
\frac{1}{\ell _{{\rm sf}}^{2}}=\left[ -\frac{e^{2}}{\tau _{{\rm sf}}^{0}} \frac{\partial
n_{s}^{0}}{\partial \mu }\right] \sum_{s}\left[ \frac{1}{\sigma _{s}}+\left( \frac{1}{\sigma
_{{\rm sf,s}}}+\frac{1}{\sigma _{D}\alpha _{s}}\right) \left( 1+\alpha _{s}\right) \right]
.\qquad
\end{equation}
This shows that the intrinsic spin relaxation length is decreased by the spin drag and angle
dependent spin flip effects.

Similarly, we get that the following weighted sum of electro chemical potentials must vanish,
\begin{equation} \nabla
^{2}(c_{-s}\bar{\mu}_{s}+c_{s}\bar{\mu}_{-s})=0,
\end{equation}
where
\begin{equation}
c_{s}=\frac{\partial n_{s}^{0}}{\partial \mu }\frac{1}{\sigma _{s}}+\left(
\frac{1}{\sigma _{{\rm sf,s}}}+\frac{1}{\sigma _{D}\alpha _{s}}\right)
\left( \frac{\partial n_{s}^{0}}{\partial \mu }+\alpha _{s}\frac{\partial
n_{-s}^{0}}{\partial \mu }\right) .
\end{equation}

Below we derive Eqs.~(\ref{cont.eq}),(\ref{ohmslaw}) and estimate the spin
drag contributions. For the two-dimensional case, we have also performed the
integration of Eq.~(\ref{rhoD}) numerically.

\section{Boltzmann equation for colinear spin transport}

We base our analysis on the Boltzmann equation for transport through a system with lifted spin
degeneracy. We take the current to run in the $x$-direction and denote the non-equilibrium
distribution function $f_{s} ({\bf k})$ and the equilibrium Fermi--Dirac distribution
function $f_{0}$
\begin{equation}
f_{0}(\varepsilon _{ks})=\frac{1}{e^{\beta \left( \varepsilon _{ks}-\mu
_{0}\right) }+1},
\end{equation}
where $\mu _{0}$ is the chemical potential and $\beta $ as usual the inverse temperature. The
eigenenergies are denoted $\varepsilon _{ks}$, where $s$ is the spin quantum number and ${\bf
k}$ the quantum number labeling the relevant states crossing the Fermi level. For simplicity
we assume a parabolic dispersion and we write
\begin{equation}
\varepsilon _{ks}=\frac{\hbar ^{2}k^{2}}{2m}+\varepsilon _{s}^{0},
\end{equation}
where $\epsilon_s$ is the band off-set which can be spin dependent if the material is
ferromagnetic.

The linearized Boltzmann equation then reads
\begin{equation}
v_{x}({\bf k)}\frac{\partial f_{s}({\bf k},x)}{\partial x}-\frac{eE_{x}} {\hbar
}\frac{\partial f_{0}(\varepsilon _{ks})}{\partial k_{x}}=\left( \frac{
\partial f_{s}({\bf k},x)}{\partial t}\right) _{{\rm coll.}}.  \label{BE}
\end{equation}
We take the collision integral to include elastic scattering and e-e scattering
\begin{equation}
\left( \frac{\partial f_{s}({\bf k})}{\partial t}\right) _{{\rm coll.}} =H_{0}[f_{s}]({\bf
k})+H_{{\rm sf}}[f_{s},f_{-s}]({\bf k})+H_{e-e}[f_{s},f_{s}]({\bf
k})+H_{e-e}[f_{s},f_{-s}]({\bf k}),
\end{equation}
where $H_{0}$ is the scattering from impurities (or quasielastic phonon scattering) giving
rise to a momentum relaxation
\begin{equation}
H_{0}[f_{s}]({\bf k})=-\int \frac{d{\bf k}^{\prime }}{(2\pi )^{d}}W_{s}^{0}
({\bf k,k}^{\prime
})\left[ f_{s}({\bf k})-f_{s}({\bf k}^{\prime })\right]
\delta (\varepsilon _{ks}-\varepsilon
_{k^{\prime }s}),
\end{equation}
and where $H_{sf}$ describes elastic scattering processes that flip the spin
\begin{equation}
H_{{\rm sf}}[f_{s},f_{-s}]({\bf k})=-\int \frac{d{\bf k}^{\prime }}{(2\pi )^{d}}W_{sf}({\bf
k,k}^{\prime })\left[ f_{s}({\bf k})-f_{-s}({\bf k}^{\prime })\right] \delta (\varepsilon
_{ks}-\varepsilon _{k^{\prime }-s}).
\end{equation}%
Finally, the e-e scattering is after the linearization given by\cite{smith}
\begin{eqnarray}
H_{e-e}[f_{s},f_{s^{\prime }}]({\bf k)} &=&-\frac{2\pi }{\hbar kT}\int
 \frac{ d{\bf k}^{\prime }}{(2\pi )^{d}}\int \frac{d{\bf q}}{(2\pi )^{d}}
 |U_{ss^{\prime }}({\bf
q},\varepsilon _{{\bf k}}-\varepsilon _{{\bf k}+{\bf q
}})|^{2}  \nonumber \\
&&\times \delta (\varepsilon _{{\bf k}}+\varepsilon _{{\bf k}^{\prime
}}-\varepsilon _{{\bf k+q}}-\varepsilon _{{\bf k}^{\prime }-{\bf q}})
\nonumber \\
&&\times f_{0}(\varepsilon _{ks})f_{0}(\varepsilon _{k^{\prime }s^{\prime
}}) \left[ 1-f_{0}(\varepsilon _{|{\bf k+q|}s}))\right] \left[
1-f_{0}(\varepsilon _{|{\bf k}^{\prime }{\bf -q|s}^{\prime }})\right]
\nonumber \\
&&\times \left[ \Psi _{s}({\bf k)+}\Psi _{s^{\prime }}({\bf k}^{\prime })- \Psi _{s}({\bf
k+q)}-\Psi _{s^{\prime }}({\bf k}^{\prime }-{\bf q)}\right] ,  \label{Hee}
\end{eqnarray}
\newline
where the deviation from equilibrium is expressed in the function $\Psi $ through
\begin{equation}
f_{s}({\bf r,k)=}f_{s}^{0}(k{\bf )+}\left( -\frac{\partial f_{0}(\varepsilon
_{ks})}{\partial \varepsilon _{ks}}\right) \Psi _{s}({\bf r,k).}
\label{Psidef}
\end{equation}
The interaction $U_{ss^{\prime }}$ is the Coulomb interaction between two
electrons with spin $s$ and $s^{\prime }$. It can in principle depend on the
relative direction of the spin if exchange is included. This set of integral
equations cannot be solved in general and one must either solve them
numerically (for example as in Ref.~\cite{hu96flen}) or proceed with
approximate methods.

One simplification is however possible from symmetry. Because of the cylindrical symmetry the
functions $\Psi _{s}({\bf k)}$ only depend on the angle between ${\bf k}$ and the direction
of the current, which we here choose to be in the $x$-direction. Denoting this angle by
$\theta $ we have $\cos \theta ={\bf k\cdot \hat{x}/}k$ and we can write
\begin{equation}
\Psi _{s}({\bf r,k})=\Psi _{s}(x,k,\theta ).
\end{equation}%
It is convenient to expand the distribution function in harmonics of the
angle $\theta $ as
\begin{equation}
\Psi _{s}(x,k,\theta )=\sum_{n=0}^{\infty }g_{s}^{(n)}(x,k)\cos n\theta,
\label{Psintheta}
\end{equation}
which we utilize in the next section.

\section{Spin drag without spin flip processes for $T\ll T_{F}$}

In this section we study the Boltzmann equation in the presence of e-e interaction, but in
the absence of spin flip processes, i.e. $H_{sf}=0.$ Furthermore, because a low temperature
expansion allows for a solution of the Boltzmann equation, we start by examining this limit
and later we discuss the validity of this solution even at elevated temperatures. It turns
out that the solution in the low temperature regime corresponds to a SFD distribution.

In the low temperature limit we see from Eq.~(\ref{BE}) that the second term
on the left hand side (the driving term) restricts $\varepsilon _{{\bf k}s}$
to lie close to the Fermi level, such that the deviation $\Psi _{s}({\bf k})$
need only be evaluate at $k_{F}$. This is therefore also true for the
distribution function in the elastic collision term, $H_{0}$. Due to the
Pauli principle this will also be the case for the $\Psi ^{\prime }$s in the
e-e collision integrals, which is seen as follows. Using standard tricks
(see e.g. \cite{jauhosmith93}), we rewrite the e-e collision term as%
\begin{eqnarray}
H_{e-e}[f_{s},f_{s^{\prime }}]({\bf k)} &=&-\frac{2\pi }{\hbar }
 \int_{-\infty }^{\infty
}\frac{d\omega }{2\pi }\int \frac{d{\bf k}^{\prime }}{(2\pi )^{d}}\int
\frac{d{\bf q}}{(2\pi
)^{d}}|U_{ss^{\prime }}({\bf q}
,\varepsilon _{k}-\varepsilon _{|{\bf k}+{\bf q|}})|^{2}  \nonumber \\
&&\times \frac{1}{kT\sinh ^{2}(\hbar \omega /k_{B}T)}
\mathop{\rm Im}%
\chi _{s}({\bf k},{\bf q};\omega )%
\mathop{\rm Im}
\chi _{s^{\prime }}({\bf k}^{\prime },-{\bf q};\omega )  \nonumber \\
&&\times \left[ \Psi _{s}({\bf k)+}\Psi _{s^{\prime }}({\bf k}^{\prime })
-\Psi _{s}({\bf
k+q)-}\Psi _{s^{\prime }}({\bf k}^{\prime }-{\bf q)}\right] ,  \label{heerw}
\end{eqnarray}
where
\begin{equation}
\mathop{\rm Im} \chi _{s}({\bf k},{\bf q};\omega )=\pi \left[ f_{0}
(\varepsilon _{|{\bf k+q}
|s})-f_{0}(\varepsilon _{ks})\right] \delta
\left( \varepsilon _{|{\bf k+q} |s}-\varepsilon
_{ks}-\hbar \omega \right) .  \label{ImXkqdef}
\end{equation}%
Now, at low temperatures the factor $1/\sinh ^{2}$ restricts the $\omega$-integral to small
$\omega $ of order $kT$, and hence $\varepsilon _{|{\bf k+q}|s}$ in (\ref{ImXkqdef}) deviates
from $\varepsilon _{ks}$ by an amount of order $kT$ and we expand $\mathop{\rm Im}\chi _{s}$
as
\begin{equation}
\mathop{\rm Im} \chi _{s}({\bf k},{\bf q};\omega )\approx \pi \hbar \omega \left(
-\frac{\partial f_{0}(\varepsilon _{ks})}
{\partial \varepsilon _{ks}}\right) \delta \left(
\varepsilon _{|{\bf k+q|}s}-\varepsilon _{ks}-\hbar \omega \right) .
\end{equation}%
From this we conclude that both $\varepsilon _{ks}$ and $\varepsilon _{k^{\prime }s^{\prime
}}$ (and hence also $\varepsilon _{|{\bf k+q}|s}$ and $\varepsilon _{|{\bf k}^{\prime }-{\bf
q}|s^{\prime }}$) are within a shell of order $k_{B}T$ from the Fermi level. To leading order
in $kT/\varepsilon _{F}$ we can therefore neglect the dependence on $k$ and keep only the
angular dependence of $\Psi _{s}$. Therefore in the following we replace
\begin{equation}
\Psi _{s}({\bf k})\approx \Psi _{s}(k_{Fs},\theta ),
\end{equation}
where $k_{Fs}$ is the Fermi wave vector for the spin direction $s$.

Now we expand the function $\Psi $ in harmonics of the angle $\theta $ as in
Eq.~(\ref{Psintheta}). Inserting (\ref{Psidef}) and (\ref{Psintheta}) into
the Boltzmann equation gives for the left hand side

\begin{equation}
\frac{\hbar k_{sx}}{m}\left( -\frac{\partial f_{0}(\varepsilon _{ks})}{
\partial \varepsilon _{ks}}\right) \left( \frac{\partial \Psi _{s}}{\partial
x}-eE_{x}\right) =\sum_{n}\frac{\hbar k_{s}\cos \theta }{m}\left( -\frac{
\partial f_{0}(\varepsilon _{ks})}{\partial \varepsilon _{ks}}\right) \left(
\cos n\theta \frac{\partial g_{s}^{(n)}}{\partial x}-\frac{eE}{\hbar }\right)
\label{lhs}
\end{equation}
and for the right hand side we have two terms. The first one is the spin
conserving impurity scattering term which becomes%
\begin{equation}
H_{0}[f_{s}]=-\sum_{n}\cos \left( n\theta \right) g_{s}^{(n)}\frac{1}
 {\tau _{{\rm tr}}^{n}}\left( -\frac{\partial f_{0}
(\varepsilon _{ks})}{\partial \varepsilon _{ks}}\right) ,
\end{equation}
where we defined the transport times of order $n$
\begin{equation}
\frac{1}{\tau _{{\rm tr},s}^{n}}=\int \frac{d{\bf k}^{\prime }}{(2\pi )^{d}} W_{s}^{0}({\bf
k,k}^{\prime })\left[ 1-\cos n\theta _{{\bf kk}^{\prime }} \right] \delta (\varepsilon
_{ks}-\varepsilon _{k^{\prime }s}),
\end{equation}%
and where $\theta _{{\bf kk}^{\prime }}$ is the angle between ${\bf k}$ and ${\bf k}^{\prime
}$. The second term is the one with the e-e scattering. When the expansion ~(\ref{Psintheta})
is inserted into the e-e interaction terms
different $n$ do not couple, see for example the derivation in Ref.~\cite%
{hu96flen}. The trick is to write for example the angle of ${\bf k}^{\prime }-{\bf q}$ as
$\cos n\theta _{{\bf k}^{\prime }-{\bf q,x}}=\cos n\left( \theta _{{\bf k}^{\prime }-{\bf
q,k}}+\theta _{{\bf k},{\bf x}}\right) =\cos n\theta _{{\bf k}^{\prime }-{\bf q,k}}\cos
n\theta _{{\bf k},{\bf x}}-\sin n\theta _{{\bf k}^{\prime }-{\bf q,k}}\sin n\theta _{{\bf
k},{\bf x}}$ and note that the $\sin $ terms vanish due to symmetry. Therefore we can express
the e-e collision term as
\begin{equation}
H_{e-e}[f_{s},f_{s^{\prime }}](k,\theta {\bf )=}\sum_{n}\cos \left( n\theta
\right) \left( g_{s}^{(n)}J^{(n)}+g_{s^{\prime }}^{(n)}I^{(n)}\right) ,
\end{equation}
where $J^{(n)}$ corresponds to the first and third term in Eq.~(\ref{heerw})
while $I^{(n)}$ corresponds to the second and fourth term.

Now a set of equations for the coefficients $g_{s}^{(n)}$ can be extracted by multiplying the
Boltzmann equation with $\cos n^{\prime }\theta $ and integrating over $\theta $, while using
that $\int d\theta \cos n\theta \cos n^{\prime }\theta \propto \delta _{nn^{\prime }}$. The
left hand side of Eq.~(\ref{lhs}) is expanded in harmonics using that $\cos \theta \cos
n\theta = \frac{1}{2}\left[ \cos \theta (n+1)+\cos \theta (n-1)\right]$. We find the
following set of equations
\begin{mathletters}
\label{dgn}
\begin{eqnarray}
\frac{\partial g_{s}^{(1)}}{\partial x} &=&0,  \label{dg1} \\
\frac{\hbar k}{2m}\frac{\partial }{\partial x}\left( g_{s}^{(0)}+\frac{g_{s}^{(2)}}{2}+e\phi
\right) \eta _{s} &=&\left( g_{s}^{(1)}\frac{\eta _{s}}{\tau _{{\rm
tr}}^{1}}+g_{s}^{(1)}\left( J^{(1)}+I^{(1)}\right)
+g_{-s}^{(1)}I^{(1)}\right) ,  \label{dg2} \\
\frac{1}{2}\frac{\hbar k}{2m}\frac{\partial }{\partial x}\left(
g_{s}^{(n-1)}+g_{s}^{(n+1)}\right) \eta _{s} &=&\left( g_{s}^{(n)}\frac{\eta _{s}}{\tau
_{{\rm tr}}^{n}}+g_{s}^{(n)}\left( J^{(n)}+I^{(n)}\right) +g_{-s}^{(n)}I^{(n)}\right) ,\quad
n\geq 2 \label{dg3}
\end{eqnarray}
where
\end{mathletters}
\begin{equation}
\eta _{s}=\left( -\frac{\partial f_{0}(\varepsilon _{ks})}{\partial \varepsilon _{ks}}\right).
\end{equation}

The solution of these equation is that $g_{s}^{(n)}=0$ for $n\geq 2.$ This
is due to the fact that $\partial _{x}g_{s}^{(1)}=0$ which decouples the
equations for $n\geq 2$ from the two first. Eq.~(\ref{dg1}) expresses
current conservation within each spin species. If we include spin flip
scattering in the equation, then the equations couple because $\partial
_{x}g_{s}^{(1)}\neq 0$.

Now we note that setting $g_{s}^{(n)}=0$ for $n\geq 2$ corresponds precisely
to a linearized shifted Fermi--Dirac (SFD)\ distribution%
\begin{equation}
f_{s}^{{\rm SFD}}(k)=f_{0}(\varepsilon _{s}({\bf k}+\delta {\bf k}%
_{s})-\delta \mu _{s}){\bf ,}
\end{equation}%
from which we read off (for $\delta {\bf k}$ in the $x$-direction),%
\begin{equation}
g_{s}^{(0)}=\delta \mu _{s},\quad g_{s}^{(1)}\cos \theta =\hbar v_{x}\delta k_{s}\Rightarrow
g_{s}^{(1)}=-\frac{\hbar ^{2}k_{Fs}}{m}\delta k_{s}. \label{sfd}
\end{equation}
From Eqs.~(\ref{dgn}), one can now determine $g_{s}^{0}$ and $g_{s}^{1}$.
They correspond to the change of the local charge densities and to the local
currents, respectively. This we will see in the next section, where we use
the SFD ansatz to study the general case.

\section{Macroscopic transport equations}

Above we saw that at low temperature the exact solution of the Boltzmann equation in absence
of spin flip was a shifted Fermi--Dirac function. The same conclusion applies to the
situation of arbitrary temperatures but weak e-e scattering, because this limit corresponds
to the usual Coulomb drag regime. This is however no longer necessarily true at arbitrary e-e
scattering, when the temperature increases, or when spin flip processes are included.
Nevertheless, we shall assume in the following that the SFD distribution is a good
approximation for the exact distribution function. The argument for doing this is as follows:
the e-e interactions will drag the distribution functions towards shifted Fermi--Dirac
distributions because the inter spin channel e-e collision terms vanish for
$f_{s}=f_{s}^{{\rm SFD}}$, i.e. $H_{e-e}[f_{s}^{{\rm SFD}},f_{s}^{{\rm SFD}}]=0$. Since the
e-e scattering rate increases as $\tau _{e-e}^{-1}\approx (\varepsilon _{F}/\hbar
)(kT/\varepsilon _{F})^{2}$, increasing the temperature actually helps. Furthermore, since the
energy dependence of the elastic scattering is important in determining the actual shape of
the distribution functions, and because we do not want to into details of this sort, we will
use view the SFD distribution functions as reasonable parameterizations of the true
distribution function.

Our starting point is thus an ansatz distribution function given by
\begin{equation}
f_{s}(k)=f_{0}(\varepsilon _{ks})+\left( -\frac{\partial f_{0}
(\varepsilon _{ks})}{\partial
\varepsilon _{ks}}\right) \delta \mu _{s}(x)-\left( -\frac{\partial f_{0}
(\varepsilon
_{ks})}{\partial \varepsilon _{ks}}\right) \hbar v_{x}\delta k_{s}(x).
\end{equation}
Here $\delta \mu _{s}$ corresponds to a change of the local chemical
potential and hence also to the local density, while $\delta k_{s}$
describes a shift of the distribution function in ${\bf k}$-space and thus
gives a finite drift velocity. Inserting this into the Boltzmann equation
gives for the left hand side

\begin{equation}
\text{LHS }=v_{x}\left( -\frac{\partial f_{0}(\varepsilon _{ks})}{\partial
\varepsilon _{ks}}\right) \left( \frac{\partial }{\partial x}\left( \delta
\mu _{s}-\hbar v_{x}\delta k_{s}\right) -eE_{x}\right) ,  \label{LHS}
\end{equation}%
and for the right hand side we have three terms. The spin conserving collision term becomes
\begin{equation}
H_{0}[f_{s}]=-\hbar v_{x}\delta k_{s}\frac{1}{\tau _{{\rm tr,}s}}\left( -%
\frac{\partial f_{0}(\varepsilon _{ks})}{\partial \varepsilon _{ks}}\right) ,
\label{H0ny}
\end{equation}
where the usual transport time is
\begin{equation}
\frac{1}{\tau _{{\rm tr},s}}=\int \frac{d{\bf k}^{\prime }}{(2\pi )^{d}}%
W_{s}^{0}({\bf k,k}^{\prime })\left[ 1-\cos \theta _{{\bf kk}^{\prime }}%
\right] \delta (\varepsilon _{ks}-\varepsilon _{k^{\prime }s}).
\end{equation}
The second term on the right hand side is the spin flip scattering term which becomes
\begin{equation}
H_{{\rm sf}}[f_{s},f_{-s}]=\left( \hbar v_{x}\delta k_{s}
\frac{1}{\tau _{ {\rm
sf,tr}}}-\left( \delta \mu _{s}-\delta \mu _{-s}\right)
\frac{1}{\tau _{ {\rm sf}}^{0}}-\hbar
v_{x}\left( \delta k_{s}-\delta k_{-s}\right) \frac{1}
{ \tau _{{\rm sf}}}\right) \left(
-\frac{\partial f_{0}(\varepsilon _{ks})}{
\partial \varepsilon _{ks}}\right) ,  \label{Hsfny}
\end{equation}
where the three different spin flip scattering times are given by
\begin{mathletters}
\label{tausff}
\begin{eqnarray}
\label{tausf0} \frac{1}{\tau _{{\rm sf}}^{0}} &=&\int \frac{d{\bf k}^{\prime }}{(2\pi )^{d}}
W_{sf}({\bf k,k}^{\prime })\delta (\varepsilon _{ks}-\varepsilon _{k^{\prime }-s}),\\
\frac{1}{\tau _{{\rm sf,tr}}} &=&\int \frac{d{\bf k}^{\prime }}{(2\pi )^{d}} W_{sf}({\bf
k,k}^{\prime })\left[ 1-\cos \theta _{{\bf kk}^{\prime }}\right]
\delta (\varepsilon _{ks}-\varepsilon _{k^{\prime }-s}),  \label{tausftr} \\
\frac{1}{\tau _{{\rm sf}}} &=&\int \frac{d{\bf k}^{\prime }}{(2\pi )^{d}} W_{sf}({\bf
k,k}^{\prime })\cos \theta _{{\bf kk}^{\prime }}\delta (\varepsilon _{ks}-\varepsilon
_{k^{\prime }-s}).  \label{tausf}
\end{eqnarray}
\end{mathletters}
Finally, the e-e scattering is given by $\sum_{s^{\prime
}}H_{e-e}[f_{s},f_{s^{\prime }}]$. But in accordance with detailed balance
the e-e scattering between two identical Fermi--Dirac distributions is zero,
$H_{e-e}[f_{s},f_{s}]=0$, and we are left with
\begin{eqnarray}
H_{e-e}[f_{s},f_{-s}] &=&-\frac{\hbar ^{2}}{m}\int_{-\infty }^{\infty }\frac{%
d\omega }{2\pi }\int \frac{d{\bf k}^{\prime }}{(2\pi )^{d}}\int \frac{d{\bf q%
}}{(2\pi )^{d}}|U_{s,-s}({\bf q},\varepsilon _{k}-\varepsilon _{|{\bf k}+%
{\bf q|}})|^{2}  \nonumber \\
&&\times \frac{1}{kT\sinh ^{2}(\hbar \omega /k_{B}T)} \mathop{\rm Im}
\chi _{s}({\bf k},{\bf q};\omega )%
\mathop{\rm Im}
\chi _{-s}({\bf k}^{\prime },-{\bf q};\omega )  \nonumber \\
&&\times q_{x}\left[ \delta k_{s}-\delta k_{-s}\right] .  \label{Heeny}
\end{eqnarray}
The final form of the Boltzmann equation is thus
\begin{equation}
(\text{\ref{LHS}})=(\text{\ref{H0ny}})+(\text{\ref{Hsfny}})+(\text{\ref {Heeny}}).
\label{BEny}
\end{equation}

Next we find the current and the density. They are given by
\begin{mathletters}
\label{jrho}
\begin{eqnarray}
J_{s} &\equiv &-e\int \frac{d{\bf k}}{(2\pi )^{d}}v_{x}f_{s}({\bf k)}
\nonumber  \label{Js} \\
&{\bf =}&e\int \frac{d{\bf k}}{(2\pi )^{d}}\hbar v_{x}^{2}\left( -\frac{%
\partial f_{0}(\varepsilon _{ks})}{\partial \varepsilon _{ks}}\right) \delta
k_{s}=\frac{\hbar e}{m}n_{s}^{0}\delta k_{s}, \\
\delta \rho _{s} &\equiv &-e\int \frac{d{\bf k}}{(2\pi )^{d}}\left[ f_{s}(%
{\bf k)-}f_{s}^{0}({\bf k)}\right]  \nonumber  \label{dns} \\
&=&-e\int \frac{d{\bf k}}{(2\pi )^{d}}\left( -\frac{\partial
f_{0}(\varepsilon _{ks})}{\partial \varepsilon _{ks}}\right) \delta \mu
_{s}=-e\frac{\partial n_{s}^{0}}{\partial \mu }\delta \mu _{s}.
\end{eqnarray}%
where is the density of states. We find two transport equations for current and charge
density or chemical potentials by integrating (\ref{BEny}) and also (\ref{BEny}) multiplied
by $v_{x}$ with respect to ${\bf k}$, and we arrive at Eq.~(1), where
\end{mathletters}
\begin{mathletters}
\label{sigmas}
\begin{eqnarray}
\alpha _{s} &=&\frac{n_{s}^{0}}{n_{-s}^{0}}, \\
\sigma _{s} &=&\frac{n_{s}^{0}e^{2}}{m}e\left( \frac{1}{\tau _{{\rm tr,}s}}+
\frac{1}{\tau _{{\rm sf,tr}}}\right) ^{-1}, \\
\sigma _{{\rm sf},s} &=&\frac{n_{s}^{0}e^{2}\tau _{{\rm sf}}}{m}.
\end{eqnarray}
In (\ref{cont.eq}) we have introduced the drag conductivity defined in Eq.~(\ref%
{rhoD}). In deriving the drag term, we have made use of the result obtained for usual Coulomb
drag in e.g.~Ref.~(\cite{jauhosmith93}). Furthermore, the local electro chemical
potential{\normalsize \ }has been defined as
\end{mathletters}
\begin{equation}
\bar{\mu}_{s}=\mu _{s}+e\phi ,
\end{equation}
where $\phi$ is the electrical potential.

\section{Evaluation of the spin drag resistivity}

\subsection{One dimension}

The polarization function is in one dimension at small temperatures, where
we can perform an $\omega $-expansion, given by%
\begin{equation}
{\rm \mathop{\rm Im} }\chi _{s}(q,\omega )\approx
\omega \frac{m^{2}}{4\hbar ^{3}q^{2}}\left(
- \frac{\partial f_{0}(\varepsilon _{q/2,s})}
{\partial \varepsilon _{q/2,s}} \right).
\end{equation}
Inserting this into the formula for $\sigma _{D}^{-1}$, performing the $\omega $-integration
for the case of a non-magnetic conductor, $\varepsilon _{ks}=\varepsilon _{k}$, and using
that $[f_{0}^{\prime }(\varepsilon )]^{2}\simeq (6kT)^{-1}\delta (\varepsilon -\varepsilon
_{F})$, we find
\begin{equation}
\sigma _{D}^{-1}\approx \frac{kT}{\varepsilon _{F}}\frac{\pi ^{2}k_{F}^{3}}{%
64}\frac{\hbar }{e^{2}}\frac{|U(2k_{F})|^{2}}{\varepsilon _{F}^{2}}.
\end{equation}
The spin-drag resistance is thus proportional to temperature and dependent
on the Coulomb backscattering matrix element. Clearly, this contribution can
be very large at finite temperatures. However, in strictly one dimensions,
where Fermi liquid theory is not expected to apply, the Boltzmann equation
is not a correct starting point and one should be somewhat careful about
drawing firm conclusions from this. Nevertheless, this Fermi's Golden rule
result is indicative of e-e interactions being very important for spin
transport in one dimension.

\subsection{Two dimensions}

For the two dimensional case we start by deriving a low temperature result and go on to
compare it with a full numerical integration of $\sigma _{D}^{-1}$. At small $\omega $ and
$q$ the imaginary part of the polarization function is given by
\begin{equation}
{\rm
\mathop{\rm Im}%
}\chi _{s}(q,\omega )\approx \omega \frac{m^{2}}{2\pi \hbar ^{3}q^{3}k_{F}},
\end{equation}
and the screened static Coulomb interaction is
\begin{equation}
U(q)=\frac{e^{2}}{2\pi \varepsilon _{0}\varepsilon _{r}}\frac{1}{(q+q_{TF})}.
\end{equation}
In this approximation, the ${\bf q}$-integral becomes
\begin{equation}
\int dq\,q^{3}\frac{1}{(q+q_{TF})^{2}},
\end{equation}
which is clearly not convergent and therefore we set the upper limit to be $%
2k_{F}$ , because $\mathop{\rm Im} \chi $ is
 zero for momentum exchange larger than $2k_{F}$.
With these
inputs, we arrive at the following approximate expression%
\begin{equation}
\sigma _{D}^{-1}\approx \left( \frac{kT}{\varepsilon _{F}}\right) ^{2}
\frac{\hbar
}{e^{2}}\frac{\pi ^{2}}{3}\frac{(1+\gamma )\ln \left( 1+\gamma \right) -\gamma }{\gamma
^{2}\left( 1+\gamma \right) },  \label{sigmaD2D}
\end{equation}
where $\gamma =2k_{F}/q_{TF}$ and $q_{TF}=me^{2}/2\pi \varepsilon _{0}\varepsilon _{r}\hbar
^{2}$ is the two dimensional inverse Thomas-Fermi screening length. Typically $\gamma $ is of
order 1. This means that spin drag resistivity can be equal to a fraction of the quantum
resistance and should therefore indeed be measurable for standard high mobility quantum wells.

\begin{figure}[t]
\setlength{\unitlength}{1cm}
\begin{picture}(14,7)
\put(3,1){\epsfysize=70mm\epsfbox{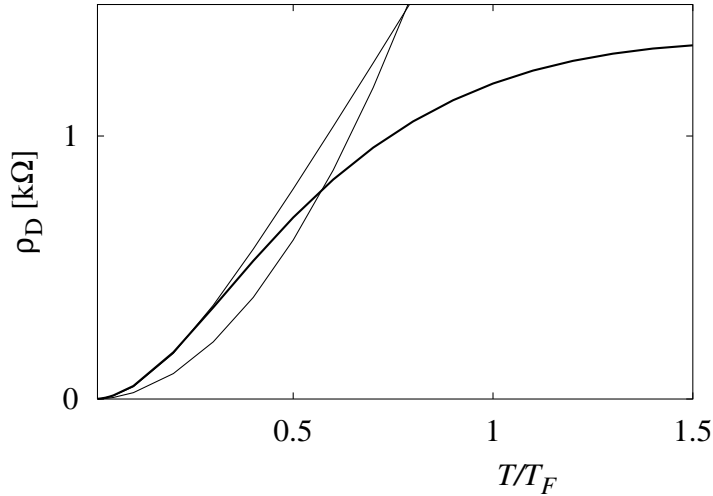}}
\end{picture}
 \caption{The spin-drag resistivity in the two-dimensional case as a
function of temperature. The full line is the numerical integration of Eq.~(\ref{rhoD}) for a
two-dimensional quantum well of thickness 10 nm, electron density 2$\times$ 10$^15$ m$^{-2}$.
We have used typical parameters for GasAs based heterostructures. The right most thin line is
the approximate expression in  Eq.~(\ref{sigmaD2D}), while the left thin line is the result
of a integrating Eq.~(\ref{rhoD}), but using the $T=0$ expression for $\chi(q,\omega)$.}
\label{fig:sigmaD2D}
\end{figure}

We have also integrated the spin drag formula numerically, see Ref.~\cite%
{flen95hu} for details. The result is shown in Fig.~\ref{fig:sigmaD2D} for realistic numbers
for a two dimensional GaAs electron gas. The integration is done using the full dynamically
screened interaction for a quantum well with finite thickness. The approximate formula
(\ref{sigmaD2D}) is seen to overestimate the spin drag effect slightly.

\section{Conclusions}

We have derived a set of transport equations for spin polarized drag which incorporate e-e
scattering. This has been done within the framework of the Boltzmann equation. First we
showed that the in the absence of spin-flip scattering and at low temperatures the exact
solution of the Boltzmann equation corresponds to two shifted Fermi--Dirac distribution
functions. Furthermore, if the interaction is weak one can use perturbation theory and arrive
at the same conclusion following the lines of argument from usual Coulomb drag. Having
observed that the shifted Fermi--Dirac distribution is correct at low temperatures or weak
e-e scattering, we go on to the general case which is solved approximately by using the SFD
as an ansatz, which allows for a solution of coupled Boltzmann equations.

The main conclusion from this is that e-e interaction introduces a spin drag term, which tend
to drag the spin currents to be equal. There a two such mechanisms namely e-e interactions,
which is temperature dependent, and angular dependent elastic spin flip scattering, which is
temperature independent. Therefore if a spin polarized current is driven through the system,
the spin drag will give rise to an additional resistivity. This resistivity increases with
temperature. We have solved for the spin drag resistivity numerically in two dimensions which
shows that it can become considerable and even exceed the ordinary impurity scattering
induced resistivity. The spin drag should thus be measurable in for example a structure
combining a two dimensional electron gas with ferromagnetic materials or for one-dimensional
systems e.g. made by nanotechnology in semiconductors or by contacting nanotubes to
ferromagnetic contacts.

\end{document}